\documentclass[prc,superscriptaddress,showpacs,twocolumn]{revtex4} 
\usepackage{graphicx,dcolumn,array,bm,amsmath}

\newcommand{\bnll}[1]{\begin{subequations}\label{#1}\begin{eqnarray}} 
\newcommand{\enll}{\end{eqnarray}\end{subequations}} 

\begin{document}
\title{Modified two-potential approach to tunneling problems}
\author{S.A. Gurvitz}
\email[]{fngurvtz@wicc.weizmann.ac.il} 
\affiliation{Department of Particle Physics,
Weizmann Institute of Science, Rehovot~76100, Israel}
\affiliation{Joint Institute for Heavy Ion Research, Oak Ridge National Laboratory, 
Oak Ridge, Tennessee 37831}
\author{P.B. Semmes}
\email[]{psemmes@tntech.edu} 
\affiliation{Physics Department, Tennessee Technological University, 
Cookeville, Tennessee 38505}
\author{W. Nazarewicz} 
\email[]{witek@utk.edu} 
\affiliation{Department of Physics and Astronomy, 
The University of Tennessee, 
Knoxville, Tennessee 37996} 
\affiliation{Physics Division, 
Oak Ridge National Laboratory, 
P.O. Box 2008, Oak Ridge, Tennessee 37831} 
\affiliation{Institute of Theoretical Physics, 
Warsaw University, 
ul. Ho\.za 69, PL-00681, Warsaw, Poland} 
\author{T. Vertse}
\email[]{vertse@tigris.unideb.hu}
\affiliation{Institute of Nuclear Research of the Hungarian Academy of Sciences \\
H-4001, Debrecen, Hungary}
\affiliation{Joint Institute for Heavy Ion Research, Oak Ridge National Laboratory, 
Oak Ridge, Tennessee 37831}


\begin{abstract}
One-body quantum tunneling to continuum is treated via the two-potential approach,
dividing the tunneling potential into  external and internal
parts. We show that corrections to this approach can be minimized
by taking the separation radius inside the interval  determined by simple
expressions. The resulting modified two-potential approach reproduces the
resonance's
energy and the width, both for  narrow and  wide
resonances.
We also demonstrate that, without losing its accuracy,
the two-potential approach can be modified to 
a form resembling the R-matrix theory, yet without any uncertainties 
related to the choice of the matching radius.   
\end{abstract}

\pacs{03.65.Nk, 03.65.Sq, 21.10.Tg, 24.30.Gd}

\maketitle

\section{Introduction}
The quantum mechanical tunneling through a 
classically forbidden
region is an ubiquitous phenomenon in physics, which has been extensively
studied since the early days of quantum mechanics. 
In 1927, 
Hund \cite{Hund} was the first to point out  the possibility of 
``barrier penetration'' between two discrete states.  In the
same year,  Nordheim  \cite{Nordheim} considered
the case of tunneling between continuum
states. Subsequently, Oppenheimer \cite{Oppy}   performed a  
calculation of the rate of
ionization of the hydrogen atom, and
Gamow, \cite{Gamow}   Gurney and  Condon \cite{Gurney} explained
alpha decay rates of radioactive nuclei
in terms of  the tunneling effect.  

While the  semi-classical treatment of tunneling 
turned out to be very successful in many applications,
the numerical calculation offers very little insight into the physical process.
In addition, the validity of the standard WKB 
formula is rather restricted. 
Other methods, although more accurate, contain various
uncertainties. For example, the results of the commonly used R-matrix
theory \cite{lane} are often  sensitive to the choice of the matching
radius \cite{humb,Arima}, and the theoretical error is difficult to estimate.

The treatment of the tunneling problem can be essentially simplified
by reducing it to two separate problems: a bound state problem  and  a
non-resonant (scattering)  state problem. This can be done consistently in the
two-potential approach (TPA) \cite{gk,g,g1} (see also
Refs.~\cite{jrb,tore}), representing the
barrier potential as a sum of the ``inner'' and the ``outer'' terms,
containing only bound and only scattering states, respectively. 
This approach
not only provides better physical insights 
than many other
approximations but it  is  also  simple and   accurate.

In this paper we propose further developments and a modification
of the TPA, and present a detailed comparison of this approach with the
results of  numerical calculations based on the Gamow-state 
(resonant-state) formalism.  The resulting analytical expressions
are  easy to interpret and they can  be straightforwardly extended to the 
non-spherical case.

The paper is organized as follows. In Sect. II,  the TPA 
is briefly described. 
Section  III deals with the quantal correction terms to the TPA.
The minimization of these terms prescribes unambiguously 
the ``window'' for the separation radius that divides the
original barrier potential into  inner and outer terms.
In this case, by considering examples of wide and
narrow nuclear resonances,  we demonstrate
that the  TPA yields results which are  practically
the same as those of the  resonant-state calculation.  
In Sec. IV,  we present a modification of the TPA.
The  resulting expressions
resemble  those of  the R-matrix theory, yet without any
uncertainties related
to the matching radius, see Sec. V.
Finally, the summary of our work is contained in Sec. VI.

\section{Two-potential approach}

Consider a   quantum well  $V(r)$ 
with  a barrier, which contains a
quasi-stationary state at the $E_{res}$.  
\begin{figure}[ht]
\begin{center}
\includegraphics[width=0.43\textwidth]{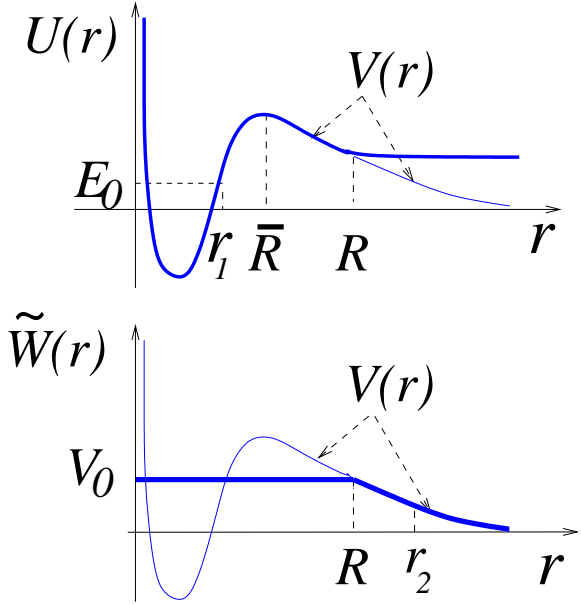}
\caption{The ``inner'' ($U$; top) and the ``outer'' ($\tilde W$; bottom)
parts of the potential $V(r)$ defined as in Eqs.~(\protect\ref{a0})
and (\protect\ref{awww}), respectively. The separation radius $R$
is chosen  well inside the barrier. The energy of the metastable
state is $E_0$ and $V_0$=$V(R)$. The barrier radius is denoted by
$\bar R$ and $r_{1,2}$ are the classical turning points. 
}
\label{Figure1}
\end{center}
\end{figure}
The coordinate space can be divided into two regions,
the ``inner''  region, $0<r<R$, and the ``outer'' region, $r$$>$$R$, where 
$R$ is taken
inside the barrier (see Fig.~\ref{Figure1}).  Accordingly, one can introduce
the two auxiliary potentials: the inner potential
\begin{equation}\label{a0}
U(r)=\left\{ \begin{array}{ll}
V(r) & {\mbox {for}}~~ r\le R \\
V(R)=V_0 & {\mbox {for}}~~ r> R
\end{array}\right.
\end{equation}
and the outer potential
\begin{equation}\label{awww}
\tilde W(r)=\left\{ \begin{array}{ll}
V_0 & {\mbox {for}}~~ r\le R \\
V(r) & {\mbox {for}}~~ r> R
\end{array}\right. .
\end{equation} 
The inner potential contains a  bound state, $\Phi_0(r)$ ($E$=$E_0$), 
representing an eigenstate of the ``inner'' Hamiltonian
$H_0=K+U(r)$, where  $K=-\nabla^2/2m$ is the kinetic energy term
($\hbar$=1).  One can  demonstrate \cite{g} that the energy and the width
of the quasi-stationary state,  associated  with  the
complex-energy poles of the  {\em total} Green's function
$G(E)=(E-K-V)^{-1}$, are obtained from the following equation 
\begin{equation} 
E=E_0+\langle\Phi_0|W|\Phi_0\rangle+\langle\Phi_0|W\tilde G(E)W|\Phi_0\rangle\, .
\label{b7}
\end{equation}
Here $W(r)=\tilde W(r)-V_0$, and the Green's function
$\tilde G$ is given by
\begin{equation}
\tilde G(E)=G_0(E)\left[1+\tilde W\tilde G(E)\right],
\label{b5}
\end{equation}
where 
\begin{equation} 
G_0(E)=\frac{1-\Lambda}{E+U_0-K-U},\;\;\;\; 
\Lambda =|\Phi_0\rangle\langle\Phi_0| .
\label{b6}
\end{equation}
The resonance
energy $E_{\rm res}=\Re(E)$ and the width
$\Gamma=-2\Im(E)$ of a quasi-stationary state obtained from Eq.~(\ref{b7})
are independent of the choice of the separation radius $R$.

Equation~(\ref{b7}) can be solved iteratively by using the standard Born series 
for the Green's function $\tilde G$, i.e. by expanding  $\tilde G$
in powers of $G_0$. Yet, the corresponding expansion for the
quasi-stationary state energy  converges very slowly. 
For that reason, we proposed \cite{g} a more efficient expansion 
scheme in which $\tilde G$ is expanded
in powers of the Green's function $G_{\tilde W}(E)=(E-K-\tilde W)^{-1}$,
corresponding to the outer potential $\tilde W$. 
From Eq.~(\ref{b5})  it immediately follows that
\begin{equation}  
\tilde G=G_{\tilde W}+G_{\tilde W}\left(U-U_0\right)\tilde G
-\tilde G_{\tilde W}\Lambda 
\left(1+\tilde W\tilde G \right)\, ,
\label{b9}
\end{equation}
Iterating Eq.~(\ref{b9}) in powers of $G_{\tilde W}$
and then substituting the result into (\ref{b7}),
one finds the desirable perturbative expansion for the  energy
and the width of the resonance. By truncating this series,
one  obtains the following first-order
relation valid  for the isolated metastable state:
\begin{equation} 
E=E_0+\langle\Phi_0|W|\Phi_0\rangle 
+\langle\Phi_0|WG_{\tilde W}(E)W|\Phi_0\rangle.
\label{b11}
\end{equation}
The above equation can be solved  iteratively
for $E=E_r-i\Gamma/2$
by assuming that the energy shift $\Delta =E_r-E_0$ and the width 
$\Gamma$ are small  compared to $E_0$ and $V_0-E_0$.
In such a  case,  one can put  $G_{\tilde W}(E)\approx G_{\tilde W}(E_0)$,
thus reducing Eq.~(\ref{b11}) to
\begin{equation} 
E=E_0+\langle\Phi_0|W|\Phi_0\rangle
+\langle\Phi_0|WG_{\tilde W}(E_0)W|\Phi_0\rangle. 
\label{bb11}
\end{equation}
By using 
the Schr\"odinger equation for $G_{\tilde W}$
one finally obtains the TPA expressions \cite{gk,g} for 
the width $\Gamma$ and for the energy shift   
$\Delta =E_{\rm res}-E_0$ of the quasi-stationary state:
\begin{eqnarray}
\strut\hspace*{-2.0em}
\Gamma 
\!\!&=&\! 
\frac{1}{mk}
\left [\Phi_0(R)\chi'_{k}(R)-\Phi'_0(R)\chi_{k}(R)\right ]^2,
\label{b17}\\
\strut\hspace*{-2.0em}\Delta 
\!\!&=&\! 
-\frac{\Phi^2_0(R)}{2mk}
\left [\alpha\chi_{k}(R)+\chi'_{k}(R)\right]
\left [\alpha\tilde\chi_{k}(R)+\tilde\chi'_{k}(R)\right],
\label{b18}
\end{eqnarray}  
where $k=\sqrt{2mE_0}$, 
$\alpha =\sqrt{2m(V(R)-E_0)}$,
$\tilde\chi_{k}(r)=\Re({\chi_{k}^{(+)}(r)})$, and  
$\chi_{k}$ ($\chi_{k}^{(+)}$) stands for 
the 
irregular (outgoing) solution of the Schr\"odinger equation
for the outer potential.

It follows from Eqs.~(\ref{b17}) and  (\ref{b18}) that both   
$\Gamma$ and $\Delta$ are given in terms of 
{\em bound} and {\em  scattering} state wave functions. 
Thus TPA essentially simplifies the treatment of tunneling,
because  the standard approximation schemes
can be used for evaluation of $\Phi_0$ and $\chi_{k}$. 
For instance, by applying the semi-classical approximation, one obtains 
the improved Gamow formula for $\Gamma$ \cite{gk,g}, which is useful  
for different applications \cite{buck,nazar}. In particular, an extension of
Eqs.~(\ref{b17}) and  (\ref{b18}) to the multi-dimensional case
can be found in Ref.~\cite{g1}. 

\section{Corrections to TPA and the choice of the separation radius}

The accuracy of Eqs.~(\ref{b17}) and  (\ref{b18})
can be determined by evaluating the leading correction terms.
There are two types of corrections to TPA: (a) those due to the replacement
of $\tilde G$ by $G_{\tilde W}$ in Eq.~(\ref{b7}) leading to Eq.~(\ref{b11}), 
and (b) those due to the replacement of 
$G_{\tilde W}(E)$ by $G_{\tilde W}(E_0)$ in Eq.~(\ref{b11})
leading to Eqs.~(\ref{b17}), (\ref{b18}).
The  correction terms of the first type, $(\Delta\Gamma )_1$, 
can be obtained by iterating  (\ref{b9}). One 
finds from the first iteration \cite{g}:
\begin{equation}
{(\Delta\Gamma)_1 \over \Gamma}\simeq {mV'(R)\over 16[2m(V_0-E_0)]^{3/2}}.
\label{d3}
\end{equation}  

Equation (\ref{d3}) might suggest that an optimal choice of the separation
radius corresponds to $V'(R)=0$, i.e.,  the maximum of $V(r)$. However,  it 
has been demonstrated numerically \cite{nazar,talou}
that  if the top of the barrier is close to
the   closing potential,
such a choice is not optimal,  since 
in this case Eqs.~(\ref{b17}) and  (\ref{b18}) become less
accurate. The reason is that  the energy shift
$\Delta$ becomes  appreciable  so that 
(\ref{b11}) cannot be replaced by (\ref{bb11}).

\begin{figure}[ht]
\begin{center}
\includegraphics[width=0.40\textwidth]{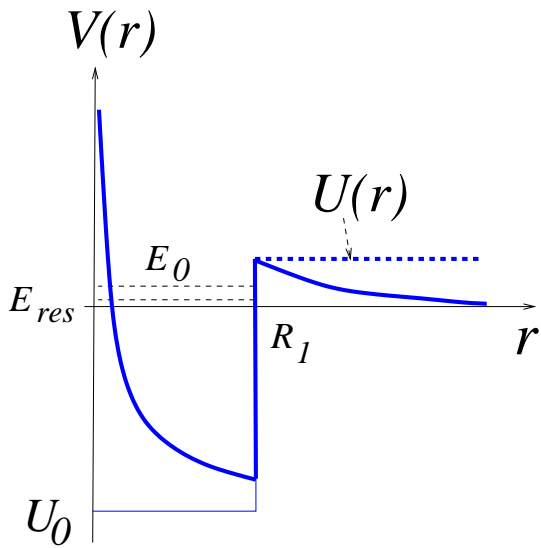}
\caption{Metastable state in the square-well potential
of Ref.~\protect\cite{erich}.
The potential parameters  ($U_0$=-51.6 MeV and $R_1$=6 fm)
correspond to a P-wave resonance at $E_{\rm res}=1$ keV  and 
$\Gamma$ =55.9 eV. 
The bound state  in
the potential $U(r)$ lies at  $E_0$=275 keV. The top of the 
centrifugal barrier is at
$V(R_1)$=1.16 MeV. 
}
\label{Figure2}
\end{center}
\end{figure}

This can be illustrated by considering 
a square-well potential discussed in Ref.~\cite{erich}: 
$V(r)=l(l+1)/2mr^2-U_0$ for $r<R_1$, 
and $V(r)=l(l+1)/2mr^2$ for $r\ge R_1$, where
the  top of the barrier coincides with the closing potential
(see Fig.~\ref{Figure2}).
For a P-wave resonance, the numerical calculation 
gives $E_{\rm res}=1$ keV and  
$\Gamma =55.9$ eV \cite{erich}. Now we apply the TPA by taking 
the separation radius at the boundary, $R=R_1$.
The corresponding inner potential 
$U(r)$  has a bound state at   
$E_0$=275~keV. However, the corresponding energy shift,
$\Delta$=$-$300 keV,  is 
of the same order of magnitude as the energy $E_0$ of the bound state.
Consequently, 
the replacement of $G_{\tilde W}(E)$ by
$G_{\tilde W}(E_0)$ in  (\ref{b11}) leads to 
large corrections to the resonance energy and the width  so that
Eqs.~(\ref{b17}) and  (\ref{b18}) cannot be used.

Let us estimate the  correction term
$(\Delta\Gamma )_2=\Gamma (E_0+\Delta)-\Gamma (E_0)$ 
due to such a replacement.
One can use the semi-classical Gamow formula, $\Gamma\propto \exp
(-2\int_{r_1}^{r_2}|p(r)|dr$, with  $r_{1}$ and $r_{2}$  being the inner and
outer classical turning points, respectively,
and $|p(r)|=\sqrt{2m[V(r)-E_0]}$.
Approximating $V(r)$ for $r_1<r<r_2$
by the inverted harmonic oscillator,  one obtains  
\begin{equation}
{(\Delta\Gamma)_2\over \Gamma}\simeq \Delta\int_{r_1}^{r_2}
{\sqrt{2m}\over
\sqrt{V(r)-E_0}}dr\approx \pi{\sqrt{2m} (r_2-r_1) 
\over 2\sqrt{\bar V-E_0}}\Delta\ ,  
\label{d4}
\end{equation} 
where $\bar V $= max $V(r)$ and   $\Delta$ is
is given by Eq.~(\ref{b18}).

Thus,  in order to reduce the correction term $(\Delta\Gamma)_2$,
one needs to minimize  the energy shift $\Delta$. It follows from
(\ref{b18}) that $\Delta$ can be strongly suppressed by
taking the separation radius $R$ deeply inside the barrier.
Indeed, Eq.~(\ref{b18}) contains a product of regular and
irregular wave functions, 
which do not vary considerably under the barrier.
However,  the  factor $|\Phi_0(R)|^2$  decays exponentially
with $R$. Therefore, by taking the separation radius $R$
far away from the boundary, $R\gg r_1$, one finds that $\Delta\to 0$,
and  $E_0\to E_{\rm res}$. As a result, (\ref{b11}) can be replaced
by  (\ref{bb11}), leading to Eqs.~(\ref{b17}) and (\ref{b18}).

To illustrate this point, let us again consider the example 
of a P-wave resonance discussed above.
By taking  the separation radius $R>R_1$, we readily 
find that $\Delta\to 0$
as  $R-R_1$ increases, and $E_0\to E_{\rm res}$= 1 keV. The resulting value of
$\Gamma$=55.3 eV  is very close to 
the Gamow-state value of the resonance width.  
The separation 
radius $R$ cannot be  chosen too close to the outer classical turning point 
since in such a case  $V(R)\to E_0$ and the correction term (\ref{d3})
becomes important. In fact, Eqs.~(\ref{d3}) and (\ref{d4})  define the lower
and upper limits of  $R$.

In the following, we discuss  the TPA results  for  proton and neutron 
resonances in the realistic average nuclear potential. 
The approximate TPA expressions 
are     compared to the   resonant
states results obtained    using the GAMOW code \cite{gamow}.

\subsection{Comparison with  Gamow-state calculations}

Consider single-nucleon resonances in the Woods-Saxon (WS) potential 
$V_{WS}(r)$ represented by a sum of central,  
spin-orbit, centrifugal, and Coulomb 
terms. Here, we apply the parametrization
of Ref.~\cite{nazar}, namely:
$R_0$=1.17A$^{1/3}$ fm, $a=0.75$ fm for the central term, and
$U^{so}_0=0.2 U_0$  and  $R^{so}_0=1.01$A$^{1/3}$ for
the spin-orbit potential.
We  calculate $\Delta$ and $\Gamma$ according to
Eqs.~(\ref{b17}) and (\ref{b18})  by  varying the separation radius $R$ 
inside the barrier, starting with the barrier radius,
$\bar R$, corresponding to the maximum of $V(r)$.

\begin{table}
\caption{
TPA calculations for the $0h_{11/2}$ and
$2s_{1/2}$ proton resonances with energy
$E_{\rm res}$=1.5\,MeV  in a WS potential. The calculated
widths $\Gamma_{\rm{TPA}}$
and the corresponding corrections (\protect\ref{d3}) and (\protect\ref{d4})
are shown relative to $\Gamma_{\rm TPA}$
for several values of the separation radius $R$. The 
actual accuracy of the TPA, $\Delta\Gamma/\Gamma$, is given
in the last column.
If the correction 
is marked zero, it means that it is below
0.1\%.
}\label{TableI}
\begin{ruledtabular} 
\begin{tabular}{|c|c|c|c|c|c|}
$R-\bar R$&$\Delta$ &$\Gamma_{\rm{TPA}}$ 
&$\underline{(\Delta\Gamma)_1}$&$\underline{(\Delta\Gamma)_2}$
&$\underline{(\Delta\Gamma)} $\\
(fm) & (keV) & (MeV) &$\Gamma_{TPA}$ &$\Gamma_{TPA}$ &$\Gamma$ \\
\hline
\multicolumn{6}{|c|}{$0h_{11/2}$ Gamow
state:
$E_{\rm res}$=1.5\,MeV, $\Gamma$ =4.918 E-18 MeV}\\
\hline
0&-1.9&4.931 E-18&0&-1\%&-0.3\% \\
1.59&-0.27&4.919 E-18&0&-0.15\%&0\\
4.28&-5.0 E-3&4.919 E-18&0&0&0\\
\hline
\multicolumn{6}{|c|}{$2s_{1/2}$ Gamow state:
$E_{\rm res}$=1.5\,MeV, $\Gamma$ =6.695 E-14 MeV}\\
\hline
0&-3.3&6.727 E-14&0&-2.3\%&-0.5\% \\
3.48&-0.11&6.709 E-14&0&-0.1\%&-0.2\%\\
8.05&-1.2 E-3&6.746 E-14&-0.7\%&0&-0.7\%
\end{tabular}
\end{ruledtabular} 
\end{table}
We begin with the high-$\ell$ narrow proton resonance  0h$_{11/2}$. 
The parameters of the  WS potential are appropriate for
$^{147}$Tm, which is a proton emitting nucleus.
The potential depth,  $U_0$=$-$61.8823 MeV, was adjusted to the
 energy $E_{\rm res}$=1.5 MeV.     
The resulting barrier radius is $\bar R$=8.54\,fm 
($V(\bar R)$=17.44 MeV) and the  inner and outer
turning points are $r_1$=6.33\,fm and $r_2$=71.15\,fm,
respectively. Since $E_{\rm res}\ll V(\bar R)$, the 
calculated 0h$_{11/2}$ resonant state has a very small width,
$\Gamma$=4.918\,10$^{-18}$\,MeV. 
The results of TPA calculations are shown in Table~I for different values of 
$R\ge\bar R$, together with  the corresponding correction
terms to TPA: $(\Delta\Gamma)_{1}/\Gamma_{\rm TPA}$ (\ref{d3}) and 
$(\Delta\Gamma)_{2}/\Gamma_{\rm TPA}$ (\ref{d4}).
Table\,I also displays
the actual accuracy of the TPA,
$(\Delta \Gamma)/\Gamma$, where $(\Delta\Gamma) =\Gamma -\Gamma_{\rm TPA}$.
Since $\bar R > r_1$, the energy shift $\Delta$
is small for $R=\bar R$. Therefore, the results of TPA
are in a good agreement with the resonant-state  calculations
already for $R=\bar R$. 
Next we consider the  low-$\ell$, broader 2s$_{1/2}$
resonance   at $E_{\rm res}$=1.5\,MeV, which is considerably closer to the  
top of the barrier $V(\bar R)=9.43$ MeV ($\bar R$=9.34\,fm). 
As  shown 
in Table\,I, also in this case 
$\Gamma_{\rm{TPA}}$ nicely agrees with the  numerical result, and 
the accuracy
of the TPA is well estimated by Eqs.~(\ref{d3}) and (\ref{d4}).

\begin{table}
\caption{
Same as in Table~\protect\ref{TableI},
except for the   neutron $0i_{13/2}$ and $1f_{5/2}$ resonances. 
}
\begin{ruledtabular} 
\begin{tabular}{|c|c|c|c|c|c|}
$R-\bar R$&$\Delta$ &$\Gamma_{\rm{TPA}}$ 
&$\underline{(\Delta\Gamma)_1}$&$\underline{(\Delta\Gamma)_2}$
&$\underline{(\Delta\Gamma )}$\\
(fm) & (keV) & (MeV) &$\Gamma_{TPA}$ &$\Gamma_{TPA}$ &$\Gamma$ \\
\hline
\multicolumn{6}{|c|}{$0i_{13/2}$ Gamow
state:
$E_{\rm res}$=1\,MeV, $\Gamma$ =1.834 E-6 MeV}\\
\hline
0&-11.9&1.869 E-6&0&-3\%&-1.9\% \\
3.45&-0.47&1.844 E-6&-1.2\%&-0.1\%&-0.5\%\\
5.96&-6.1 E-2&1.847 E-6&-1.5\%&0&-0.7\%\\
\hline
\multicolumn{6}{|c|}{$1f_{5/2}$ Gamow state:
$E_{\rm res}$=1\,MeV, $\Gamma$ =9.271 E-2 MeV}\\
\hline
0&-109&1.227 E-1&0&-25\%&-32\% \\
2.05&-51.4&1.089 E-1&-10.1\%&-12\%&-17\% \\
4.05&-7.2&9.876 E-2&-12\%&-1.7\%&-6.5\%\\
4.57&+2.7 &9.408 E-2&-17\%&+0.6\%&-1.5\%
\end{tabular}
\end{ruledtabular} 
\label{TableII}
\end{table}
Table\,II displays the  TPA results for  the
0i$_{13/2}$ and 1f$_{5/2}$
neutron resonances in  $^{133}$Sn
at an energy $E_{\rm res}$=1\,MeV. Here
$\Gamma$
is much larger due to the absence of the Coulomb barrier. 
As in the proton  case, there is  very good
agreement with numerical calculations, provided that  
$R$ is taken far away from the turning points, inside the window
determined by (\ref{d3}) and (\ref{d4}), and 
the results of TPA weakly depend   on the separation radius $R$.
This suggests that the separation 
radius can be eliminated altogether  from the TPA expressions.
As demonstrated in the following section one can indeed modify
the TPA in such a way.

\section{Modified Two-Potential Approach}

\begin{figure}[ht]
\begin{center}
\leavevmode
\includegraphics[width=0.40\textwidth]{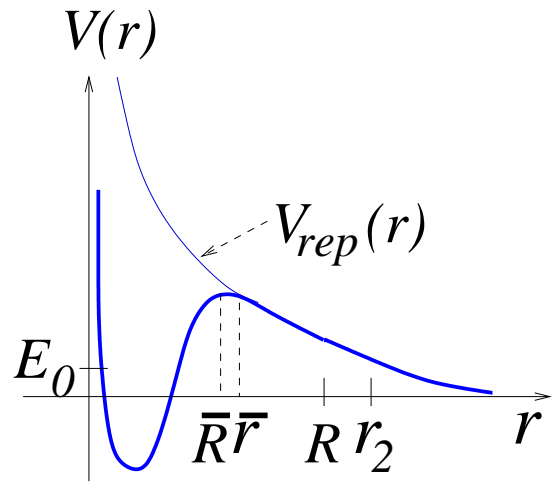}
\caption{For $r>\bar r$
the tunneling  potential $V(r)$ can be 
approximated by its repulsive part $V_{rep}(r)$. $R$ and 
$r_2$ denote the  TPA separation radius
and the outer turning
point ($V(r_2)=E_0$), respectively.  
}
\label{Figure4}
\end{center}
\end{figure}

A tunneling  potential can always be written  as a sum
of attractive and repulsive parts,
$V(r)=V_{att}(r)+V_{rep}(r)$, where 
$V_{rep}(r)$ becomes  dominant at distances beyond the barrier
radius (see Fig.~\ref{Figure4}). 
Therefore, starting with some radius $\bar r$, 
the total potential $V(r)$ can be well approximated 
by the repulsive component only. 
The value of $\bar r$ should be chosen in such a way 
that the attractive (e.g., nuclear) part can disregarded 
with a desired accuracy $\eta\ll 1$: 
\begin{equation}
|1-V_{rep}(r)/ V(r)|\le\eta~~~~~~{\mbox{for}}~r\ge \bar r\, .
\label{e1}
\end{equation}
For instance, in the case of a square-well
potential of
Fig.~\ref{Figure2},
Eq.~(\ref{e1}) is satisfied for any $\eta$ provided
that $\bar r$$>$$R_1$. In most cases, the
attractive potential 
 rapidly decreases beyond
the barrier radius, so that  $\bar r$ is  closer
to $\bar R$ than to the
separation radius $R$ in the TPA (cf. Fig.~\ref{Figure4}).

Usually,  the repulsive part $V_{rep}(r)$
is well known, as well as the   two linearly independent
(regular and irregular) solutions $F_k(r)$ and $G_k(r)$
of the corresponding Schr\"odinger
equation. For instance, if $V_{rep}(r)$
is a sum of Coulomb and centrifugal potentials, then
$F_k(r)$ and $G_k(r)$ are the standard Coulomb
functions.
This implies that any solution of the Schr\"odinger
equation with the potential $V(r)$ 
can be written for 
$r>\bar r$ as the linear combination of
$F_k(r)$ and $G_k(r)$. 

Consider  the bound-state wave function $\Phi_0(r)$ of 
the inner potential $U(r)$ of Eq.~(\ref{a0}).
Since $U(r)\simeq V_{rep}(r)$ for $\bar r\le r\le R$, $\Phi_0(r)$
can be expanded in this region  as
\begin{equation}
\Phi_0(r)=c_1G_k(r)+c_2 F_k(r)\, ,  
\label{ee3}
\end{equation}
where $k=(2mE_0)^{1/2}$. The coefficients $c_{1,2}$
and the  energy $E_0$    
are obtained from matching of the logarithmic derivatives at  
$r=\bar r$ and $r=R$:
\bnll{e3}
{c_1G'_k(R)+c_2 F'_k(R)\over c_1G_k(R)+c_2 F_k(R)}&=&-|p(R)|,\\
{c_1G'_k(\bar r)+c_2 F'_k(\bar r)\over c_1G_k(\bar r)+c_2 F_k(\bar r)}
&=&{\Phi'_0(\bar r)\over\Phi_0(\bar r)}. 
\enll
Note that $\Phi_0(r)\propto \exp [-|p(R)|(r-R)]$, for $r>R$.
Solving Eqs.~(\ref{e3}) one easily finds 
\begin{equation}
{\Phi_0'(\bar r)\over\Phi_0(\bar r)}=
{G'_k(\bar r)\over G_k(\bar r)}
\left [{1+\alpha  a_k^{-1}-f_1(1+\tilde a_k^{-1})
\over 1+\alpha a_k^{-1}-f_2(1+\tilde a_k^{-1})}\right ]\ ,
\label{e4}
\end{equation}
where $\alpha =|p(R)|$,  
$a_k={F'_k(R)/F_k(R)}$,
$\tilde a_k={G'_k(R)/G_k(R)}$, and
\begin{equation} 
f_1={G'_k(R)F'_k(\bar r)\over
F'_k(R)G'_k(\bar r)},~~~ 
f_2={G'_k(R)F_k(\bar r)\over
F'_k(R)G_k(\bar r)}. 
\end{equation}

The  wave functions 
$F_k(r)$ and $G_k(r)$ are of the same order of magnitude
in the asymptotic region, $r\gg r_2$.
However, in the classically forbidden region
the regular wave function 
exponentially decreases and the irregular one exponentially increases
with decreasing  $r$.  
Using the semi-classical approximation, one can estimate
$G_k(R)F_k(\bar r)\sim\exp[-\int_{\bar r}^R|p(r)|dr]$. 
Therefore,  the coefficients $f_{1,2}$ are of the order of
$\exp [-\alpha (R-\bar r)]$, so the
corresponding terms $f_{1,2}(1+\tilde a_k^{(-1)})$ in Eq.~(\ref{e4})
are exponentially suppressed  and can be neglected. As a result, the 
matching condition (\ref{e4}) can be written as
\begin{equation}
{\Phi'_0(\bar r)\over\Phi_0(\bar r)}=
{G'_k(\bar r)\over G_k(\bar r)}\,.
\label{e5}
\end{equation}
The above equation constitutes
 the MPTA  condition for the resonance energy $E_{\rm res}=E_0$.
In contrast to Eq.~(\ref{e4}), the relation (\ref{e5}) 
 does not exhibit any explicit $R$-dependence.
Consequently,  there is no need to evaluate the bound-state wave function
at  large values of $R$  well inside the barrier.
Note also that for a narrow resonance the
irregular wave function $G_k(r)$  
is proportional to  
the real part of the outgoing (Gamow) solution:
\begin{equation}               
\Psi^{out}_{k_{res}}(r) \propto G_{k_{res}}(r)+iF_{k_{res}}(r)=O_{k_{res}}(r),
\end{equation} 
with complex $k_{res}=k-i\gamma$ and
$k_{res}^2=2m(E_{res}-i\Gamma/2)$. Since for small
$\Gamma$ and $\gamma$ the imaginary parts of $k_{res}$
and $G_{k_{res}}(r)$
can be neglected,  $k_{res}\approx k$ and
$G_{k_{res}}(r)\approx G_{k}(r)$ (also for the
outgoing Coulomb wave function  $O_{k_{res}}(r)=O_{k}(r)$). In this case
Eq.~(\ref{e5}) represents the matching condition for the
inner (bound state) wave function with the {\em real
part} of the non-normalized outgoing
wave. If the imaginary parts are not
negligible,  Eq.~(\ref{e5}) 
can be still interpreted in terms of the standing-wave boundary 
condition at $\bar r$,   which
means that the scattering phase shift is $\pi/2$.
The requirement that the phase shift is equal to $\pi /2$ 
represents an alternative definition for the position of a
resonance in the absence of a  non-resonant phase shift.
                          
Consider now Eq.~(\ref{b17}) for the width. Since $R>\bar r$,  
the outer wave function $\chi_k(r)$ in the region $r\ge R$
can be represented by the linear combination of the regular
and irregular solutions of  $V_{rep}(r)$, and
the corresponding coefficients
in the linear combination of $F_k(r)$ and $G_k(r)$ are directly
related to the 
({\em non-resonant})
scattering phase shift for the outer potential $\tilde
W(r)$. One easily finds 
\begin{equation}
\chi_k(r)=
\cos\delta_k\ F_k(r)+\sin\delta_k\ G_k(r)~~~{\mbox{for}}~~r\ge R\, ,
\label{e8}
\end{equation}
where the phase shift $\delta_k$ is obtained from matching
of logarithmic derivatives at the separation radius $R$: 
\begin{equation}
\tan\delta_k=-{F_k(R)\over G_k(R)}\left 
({a_k-\alpha \over\tilde a_k-\alpha }\right )\ . 
\label{e9}
\end{equation}
Here we neglected the terms $\sim\exp (-2\alpha R)$.
Substituting (\ref{ee3}) and (\ref{e8})
into (\ref{b17}) and taking into account the  Wronskian relation 
between $F_k(r)$ and  $G_k(r)$ 
one obtains:
\begin{equation}\label{e10}
\Gamma=\cos^2\delta_k{k\over m}
\left [{\Phi_0(\bar r)\over G_k(\bar r)}
\right ]^2\left [1-{(\tilde a_k+\alpha )
(a_k-\alpha)\over (a_k+\alpha )(\tilde a_k-\alpha )}\right ]^2.
\end{equation}
Note that in the classically forbidden region 
$F'_k(r)/F_k(r)\approx |p(r)|$ and 
$G'_k(r)/G_k(r)\approx -|p(r)|$, so that 
the second term in brackets of (\ref{e10}) can 
be neglected. In addition, $\cos\delta_k\simeq 1$, as follows from
(\ref{e9}). As a result, one arrives at   
the following simple expression for the width:
\begin{equation}
\Gamma ={k\over m}\left [\frac{\Phi_0(\bar r)}{G_k(\bar r)}
\right ]^2.
\label{e11}
\end{equation}
Thus, similar to Eq.~(\ref{e5}) for the resonance's energy, the separation 
radius $R$ does not appear explicitly in the expression for the width.

Equations~(\ref{e5}) and (\ref{e11}) 
represent the final result of 
the modified two-potential approach (MTPA).
Despite their simple appearance, these expressions are 
very accurate. In fact,
the accuracy of the MTPA is practically the same as that
 of the  TPA since
the former was derived from
the latter by neglecting only small correction  terms of the order
of the accuracy of the TPA itself. 
For instance, for the previously
discussed case of the P-wave resonance
in the square well potential,
one finds $E_{\rm res}$=1\,keV  and $\Gamma$=55.3\,eV, i.e.  
the same result as in TPA.  In general, the accuracy of the MTPA
can be estimated  by means of  the parameter $\eta$, which
defines the lower limit for the matching radius $\bar r$. However,
one has to keep in mind that 
$\bar r$ cannot be very large 
since the derivation of Eqs.~(\ref{e5}) and (\ref{e11}) is valid only for 
$\alpha (R-\bar r)\gg 1$. Therefore, the value of  $\eta$ 
is restricted by Eq.~(\ref{d3}) in which $R$
is replaced by $\bar r$.

It is worth noting that 
an expression similar to Eq.~(\ref{e11}) 
was used in Refs.~\cite{maglione,ferreira} for calculating
partial widths for proton emission. The corresponding
formula that applies to the single channel case can be written as
\begin{equation}
\Gamma(r) ={k\over m}
\left|{\Re\left(\Psi^{out}_{k_{res}}(r)\right)\over 
O_{k}(r)}\right|^2\ ,
\label{e30}
\end{equation}
where $r$ is large. The $r$-dependence of  $\Gamma(r)$
is weak but it can be reduced if one takes  a more
appropriate expression: 
\begin{equation}
\Gamma(r) ={k\over m}
\left|{\Psi^{out}_{k_{res}}(r)\over
O_{k_{res}}(r)}\right|^2\ .
\label{e31}
\end{equation}                                           
Another expression for the width can be derived from 
the continuity relation for the resonant states~\cite{humb,barmore}:
\begin{equation}
\Gamma(r)=i{{1}\over{2m}} {{\Psi^{out\prime*}_{k_{res}}(r)\Psi^{out}_{k_{res}}(r) 
-\Psi^{out\prime}_{k_{res}}(r)\Psi^{out*}_{k_{res}}(r)}\over{\int_0^{r} 
|\Psi^{out}_{k_{res}}(r')|^2dr'}}.
\label{e32}
\end{equation}                  
This form is completely independent of $r$ in a  
wider range \cite{barmore,vertseproc} and furnishes a
value which is equal to that coming from the imaginary part
of the energy. However, for very narrow resonances,
it is difficult to calculate the imaginary part of the
energy with sufficient precision.  
The expression (\ref{e11})
derived in the MTPA
replaces  the Gamow wave function
with the (real) bound-state  wave function $\Phi_0$.  
Finally, let us emphasize that 
while Eq.(\ref{e30}) resembles the MTPA expression,  
it is based on different approximations and boundary conditions.
On the other hand there are close connections between
the MTPA and the R-matrix theory, also employing  real-energy
eigenstates, see Sec.~\ref{Rmatrix}.

\subsection{MTPA: numerical examples}

We present below in Table III  the results of the MTPA for the widths
 (\ref{e11}) of resonances discussed in Tables I and II in the context 
 of TPA. (Since
the  MTPA resonance energies  are very
close to the exact result,  they are not displayed.)
One finds that the MTPA reproduces the width almost with
the same accuracy as  the TPA, provided that the matching radius
$\bar r$ is large enough to ensure that the  contribution
from the nuclear attractive potential
is small  ($\eta\ll 1$). 
\begin{table}
\caption{Similar as in Tables~\protect\ref{TableI} and 
\protect\ref{TableII}, except for the MTPA.}
\begin{ruledtabular} 
\begin{tabular}{|c|c|c|c|}
$\bar r-\bar R$ &$\eta$&$\Gamma_{\rm{MTPA}}$ &
$\underline{(\Delta\Gamma)}$\\
(fm) & & (MeV) &$\Gamma$ \\
\hline
\multicolumn{4}{|c|}{$0h_{11/2}$ Gamow state: $E_{\rm res}$=1.5\,MeV, $\Gamma$ =4.918 E-18 MeV}\\
\hline
0.17&0.11&4.665 E-18&5\% \\
1.55&0.02&4.87 E-18&1\%\\
3.09&0.003&4.909 E-18&0.2\%\\
\hline
\multicolumn{4}{|c|}{$2s_{1/2}$ Gamow state: $E_{\rm res}$=1.5\,MeV, $\Gamma$ =6.695 E-14 MeV}\\
\hline
0.26&0.06&6.577 E-14&2\%\\
1.66&0.01&6.675 E-14&0.3\%\\
3.24&0.001&6.692 E-14&0 \\
\hline
\multicolumn{4}{|c|}{$0i_{13/2}$ Gamow state: $E_{\rm res}$=1\,MeV,
$\Gamma$ =1.834 E-6 MeV}\\
\hline
0.18&0.15&1.736 E-6&5\% \\
1.45&.04&1.814 E-6&1\%\\
2.99&.007&1.831 E-6&0.1\%\\
\hline
\multicolumn{4}{|c|}{$1f_{5/2}$ Gamow state: $E_{\rm res}$=1\,MeV, $\Gamma$ =9.271 E-2 MeV}\\
\hline
0.18&0.13&8.998 E-2&3\%\\
1.38&0.035&8.856 E-2&4\%\\
2.96&0.005&8.373 E-2&10\%
\end{tabular}
\end{ruledtabular} 
\label{TableIII}
\end{table}
It follows from Table III  that $\eta$
controls the accuracy of MTPA rather well, except for 
a broad neutron   resonance 1$f_{5/2}$ when
$\Delta\Gamma /\Gamma$  reaches $10\%$ at 
$\bar r=\bar R+2.96$. In this case,
the matching radius is quite far away from
the barrier
radius, so that
the accuracy of the MTPA is given by
Eq.~(\ref{d3})
(with $R$ replaced
by $\bar r$). This is well confirmed by Table II, which shows 
the corresponding correction term. 

\section{Comparison with the R-matrix theory}\label{Rmatrix}

It is interesting to compare the final expressions of the MTPA,
Eqs.~(\ref{e5}) and  (\ref{e11}), with the results of
the R-matrix theory \cite{lane}. In the latter method,  the 
space is divided  into  internal and external regions
by a hard sphere of the
radius $a_c$, and a complete set of the internal wave functions
$u_\lambda(r)$ is introduced,
$\int_0^{a_c}u_\lambda(r)u_{\lambda'}(r)dr=\delta_{\lambda\lambda'}$.
The internal wave functions obey  real 
boundary conditions for the logarithmic derivative:
\begin{equation}
\frac{u'_\lambda (a_c)}{u_\lambda(a_c)}
=B\, .
\label{e37}
\end{equation}
The value of $B$ determines  the R-matrix energy
eigenvalues
$E_\lambda$. It is convenient to choose B so that
one of the eigenvalues would coincide with the position of the
resonance where the value of the phase shift is equal
to $\pi/2$. This defines
the so-called ``natural'' boundary condition: 
\begin{equation}
\frac{u'_\lambda (a_c)}{u_\lambda(a_c)}
=\Re\left [
\frac{G_{k_\lambda}'(a_c)}{G_{k_\lambda}(a_c)}\right ].
\label{e7}
\end{equation}
In this case, one can apply the one-level approximation,
i.e.,  approximate the resonance with a single
Breit-Wigner term.
The corresponding width $\Gamma_\lambda$ becomes:
\begin{equation}
\Gamma_\lambda ={k\over m[ 1-g_\lambda (a_c)]}{u^2_\lambda(a_c)
\over |G_k(a_c)|^2}\, ,
\label{ee7}
\end{equation}
where 
$g_\lambda$ denotes the energy derivative of
the level shift \cite{lane,erich1}.

One finds that Eqs.~(\ref{e5}) and (\ref{e11}) of the MTPA 
formally resemble Eqs.~(\ref{e7}) and (\ref{ee7}) by choosing $a_c=\bar r$
and taking $g_\lambda (a_c)=0$.  
Yet, the inner wave function $\Phi_0(r)$ of the MTPA
is different from
the internal wave function $u_\lambda(r)$ of the R-matrix theory.
The latter is totally confined inside the inner region, whereas
$\Phi_0(r)$ is a true ``bound state'' wave function of the inner
potential $U(r)$. Therefore, their normalizations 
are different. 

The essential problem of the R-matrix theory is a proper
choice of the matching radius $a_c$, which remains  a free  parameter.
A different choice of $a_c$ does affect the
results of R-function calculations in a one-level approximation. Moreover,
there  should exist an optimal  matching radius, for which
the results of the R-matrix calculations are  close
to the exact results \cite{erich}. However,
except for some simple cases (e.g., the square well potential),
the optimal matching radius cannot be simply prescribed.
In contrast, the MTPA is not sensitive to the matching radius
$\bar r$, provided that it  is taken inside the ``window"
defined by  Eqs.~(\ref{d3}) and  (\ref{e1}).
This is an essential advantage of MTPA over the R-matrix method.
(For   critical discussion of the R-matrix expression 
for the resonance width, see Refs.~\cite{Arima,barmore,kruppaproc}.)

\section{Summary}  

This paper contains a detailed investigation  of the two-potential
approach to the one-body
tunneling problem. It has been found  that TPA  becomes
extremely accurate if the separation radius, dividing the entire space into
the inner and the outer regions, is taken deeply inside the barrier,
but not too close to the outer classical turning point. From a 
minimization of the leading correction terms,
we obtained simple expressions for the upper and lower limits 
of the TPA separation radius. 
The high accuracy
of the method  was demonstrated explicitly by a detailed comparison
with  Gamow resonances of a
realistic nuclear potential. 

Furthermore, we have found that the TPA can be further simplified 
 by taking into account the properties of regular and irregular
solutions of the Schr\"odinger equation under the barrier. The final
expressions of the  modified two-potential approach formally resemble
those of the R-matrix theory with the ``natural''  
boundary conditions. However,  the internal wave function of
the MTPA is considerably different.
In addition, contrary to the R-matrix theory,
the corresponding matching radius of the MTPA is well
defined. This makes MTPA particularly suitable  for practical applications.

\section{Acknowledgments}
One of the authors (S.G.) is very grateful to E. Vogt for inspiring 
discussions and important suggestions.  S.G. also thanks 
TRIUMF, Vancouver and JIHIR  for kind hospitality.
This work was supported in part by the U.S.~Department of Energy under Contract
Nos.~DE-FG02-96ER40963 (University of Tennessee), DE-AC05-00OR22725 with UT-Battelle,
LLC (Oak Ridge National Laboratory), and DE-FG05-87ER40361 (Joint Institute for Heavy
Ion Research).
T.V. acknowledges support from the Hungarian  OTKA Grants No. 
T029003 and T037991. The support from the University Radioactive Ion Beam consortium
(UNIRIB) is gratefully acknowledged.

\end{document}